\documentstyle[eqsecnum, epsf, aps]{revtex}

\newcommand{\f}{\phi(x)}

\newcommand{\ue}{\eta^{\mu \nu}}
\newcommand{\kf}{\tilde{\phi}(k)}

\begin{document}

\draft

\title{ On a Generalization in Quantum Theory: Is $\hbar$ Constant? }

\author{ Ronald J. Adler\thanks{adler@relgyro.stanford.edu}} 
\address{ Gravity Probe B, W.W.Hansen Experimental Physics Laboratory, \\ 
 Stanford University, Stanford, CA 94305-4085}
\author{ David I. Santiago\thanks{david@spacetime.stanford.edu}}
\address{Gravity Probe B and Department of Physics,\\ 
 Stanford University, Stanford, CA 94305-4060}

\date{\today}

\maketitle

\begin{abstract}
We here consider a generalization of the Klein-Gordon scalar wave
equation which involves a single arbitrary function. The quantization may be 
viewed as allowing  $\hbar$ to be a function of the momentum or wave vector 
rather than a constant. The generalized theory is most easily viewed in the 
wave vector space analog of the Lagrangian. We need no reference to spacetime. 
In the generalized theory the de Broglie relation between wave vector and 
momentum is generalized, as are the canonical commutation relations and the 
uncertainty principle. The generalized uncertainty principle obtained is the 
same as has been derived from string theory, or by a general consideration of 
gravitational effects during the quantum measurement process. The propagator of
the scalar field is also generalized, and an illustrative example is given in 
which it factors into the usual propagator times a "propagator form factor."
\end{abstract}


\section{ Introduction}

Early in the history of quantum mechanics de Broglie \cite{debrog} suggested 
that a wave might be associated with a particle such as an electron, and that
the wave vector $\vec{k}$ be related to the momentum $\vec{p}$ by
\begin{equation}
\vec{k} = \frac{\vec{p}}{\hbar} \label{momtm} \, ,
\end{equation}
or, in terms of the wavelength,
\begin{equation}
\lambda = \frac{2 \pi \hbar}{|\vec{p}|} \label{wvlngth0} \, .
\end{equation}
This ad hoc suggestion is widely considered as an important step leading to
the Schroedinger equation and nonrelativistic quantum mechanics.

In this work we study a generalization of this relation. The generalization may
be viewed as allowing $\hbar$ to be a function of the momentum or the wave 
vector,  or more precisely as a function of the square of the
four-momentum or wave vector; that is we take $\hbar \rightarrow \hbar g(p^2)$ 
or equivalently $\hbar \rightarrow \hbar f(k^2)$. In this paper we study this 
in the context of a generalized Klein-Gordon or scalar wave equation, which we 
study in both spacetime, where it may have an arbitrarily large number of 
derivatives, and also in wave vector space, where it contains the arbitrary 
function $f(k^2)$. We emphasize that the analysis is particularly simple and 
logically complete in wave vector space, and that spacetime need not be 
considered at all.

The generalized de Broglie relation leads to a modified canonical commutation 
relation between position and momentum, and that in turn leads to a generalized
uncertainty principle, or GUP. The GUP is the same as that obtained in string 
theory by Veneziano and other authors \cite{ven}; it has also been obtained 
from a very general consideration of gravitational effects in the quantum 
measurement process \cite{gup}. Since we obtain the GUP here without reference 
to gravity we may speculate that there is some connection between the 
generalized de Broglie relation and gravity, or that gravity is somehow 
generated by intrinsic quantum effects \cite{sak}.

Lastly we very briefly consider interaction of the scalar field with a given 
external scalar field in the context of wave vector space.  Of course the
propagator of the scalar field is generalized, and with one illustrative choice
of the function $f(k^2)$, we show that the generalized propagator is equal to 
the usual propagator times a  ``propagator form factor.'' The form factor has 
the form of the propagator of a virtual particle of Planck mass.

Some time ago Snyder \cite{sny} considered the idea of a quantized or 
``discretized'' spacetime. His ideas are equivalent to our modified de Broglie 
relation, but his  motivation was different from ours. Recently, from a 
somewhat different perspective, some of our results have been derived by Kempf
and collaborators \cite{kempf}. In particular, they obtain our results 
regarding commutators and uncertainty relations and construct a theory that is 
regular in the ultraviolet. Our philosophy and approach in the present work is 
different than theirs, and we treat interactions differently.

\section{ The scalar wave equation in spacetime }

We first recall briefly some simple properties of the scalar wave
equation in spacetime. Throughout we use units in which $c=1$ but take pains to
retain $\hbar$ explicitly. The Lagrangian density for a noninteracting or free 
scalar field is
\begin{equation}
 L= \hbar^2 \f^{*}_{, \mu} \f_{, \nu} \ue - m^2 \f^* \f \, , \ \ \ \f_{,\alpha}
\equiv \frac{\partial \f}{\partial x^{\alpha}} \, ,\label{stlag}
\end{equation}
where $\ue$ is the Lorentz metric with signature (1,-1,-1,-1), and the scalar
field and its conjugate are treated as independent variables.  From this
the Euler-Lagrange equations are
\begin{equation}
\hbar^2 \f_{, \mu \, , \nu} \ue +m^2 \f =0 \label{stkg} \, ,
\end{equation}
and similarly for the conjugate. As a solution we seek a plane wave of the form
\begin{equation}
\f = \phi_0 e^{-i k_{\beta} x^{\beta}} \label{plnwv} \, .
\end{equation}
Here the wave vector $k_{\beta}$ is a set of four real parameters. 
Substitution of this into the wave equation (\ref{stkg}) shows it is a solution
if
\begin{equation}
\hbar^2 (k_{\mu} k_{\nu} \ue) \equiv \hbar^2 k^2 = m^2 \label{dispersion}
\end{equation}
Special relativity tells us that the four-momentum squared of a particle is
equal to its mass squared, $p^2 = m^2$. Comparing this with (\ref{dispersion}) 
we naturally assume the wave vector is related to the momentum by $k_{\mu} = 
p_{\mu} / \hbar$ , which we refer to as the \underline{de Broglie relation}. 
This is not the most general solution, as we will discuss in   section IV. 
In particular the de Broglie relation implies the usual relation between the 
wavelength $\lambda = 2 \pi / |\vec{k}|$ and the magnitude  of the 
three-momentum $p= |\vec{p}|$
\begin{equation}
\lambda = \frac{2 \pi \hbar}{p} \, . \label{debrog}
\end{equation}
Since the wave vector may be identified with the derivative operator $i\partial
 / \partial x^{\alpha}$ this gives the standard operator expression for 
momentum in a position representation, and the standard commutation relation
\begin{equation}
p_{\gamma} = i \hbar \frac{\partial}{\partial x^{\gamma}} \, , \ \  [x^{\alpha}
, \, p_{\beta}] = -i\hbar \label{stcomm} \, .
\end{equation}
In a section IV we will modify and generalize the equations in this section.

\section{ The scalar wave equation in wave vector space}

It is interesting to express the action principle for fields in the space
of wave vectors, without reference to spacetime. This is an elementary
exercise in Fourier transforms, but the physical viewpoint is different,
and may allow more flexibility and  insight when seeking
modifications of present theories. We use the scalar field to illustrate
this. The action is defined as the integral of the Lagrangian in spacetime,
\begin{equation}
S = \int d^4 x L[\f, \, \f_{, \mu}] \label{staction} \, ,
\end{equation}
where the integral may be taken over all space, and time from an initial
time, $t_i$, to a final time, $t_f$. (Dependence of the Lagrangian on the 
conjugate field is implicit.) In particular we may take $t_i \rightarrow - 
\infty$ and $t_f \rightarrow \infty$ so the integral is over all spacetime. If 
we Fourier transform the field in spacetime ($x$) to wave vector space ($k$) by
\begin{equation}
\f = \int \frac{d^4k}{(2 \pi)^4} \kf e^{-ikx} \, , \label{transfrm}
\end{equation}
then the action may be expressed in terms of an analog of the Lagrangian in
wave vector space,
\begin{equation}
S= \int \frac{d^4k}{(2 \pi)^4} K[\kf, k_{\mu} \kf] \, , \ \ \
K[\kf, k_{\mu} \kf] \equiv \hbar^2 [k_{\mu} \kf^* ] [ k_{\nu} \kf ] \ue - m^2 
\kf^* \kf \, . \label{kaction}
\end{equation}

To obtain the appropriate dynamical equations in wave vector space
we extremize the action with respect to variations of the fields. This can
be done in two equivalent ways. In the first way we take the fields, $\kf$ and
$\kf^*$, and the product of the wave vector and the fields,$k_{\mu} \kf$  and  
$k_{\mu} \kf^*$, as independent variables; this may appear somewhat artificial,
but it is the analog of treating the field $\f$ and its derivatives 
$\f_{, \mu}$ as independent variables in spacetime, and is useful for formal 
manipulations. Extremizing in this way we obtain
\begin{equation}
\delta S = \int \frac{d^4k}{(2 \pi)^4} \left[ \left( \frac{\partial K}{\partial
 \kf} \delta \kf + \frac{\partial K}{\partial (k_{\mu} \kf)} \delta ( k_{\mu}
\kf) \right) + \left(  \frac{\partial K}{\partial \kf^*} \delta \kf^* + \frac{
\partial K}{\partial (k_{\mu} \kf^*)} \delta ( k_{\mu} \kf^*)\right)  \right]
=0 \, . \label{vars}
\end{equation}
Since the variations of the fields are arbitrary, and since $\delta ( k_{\mu}
\kf) = k_{\mu} \delta \kf$, this leads to an analog of the Euler-Lagrange 
equations
\begin{equation}
\frac{\partial K}{\partial\kf} + \frac{\partial K}{\partial (k_{\mu} \kf)} k_{
\mu} = 0 \, , \label{kwveq}
\end{equation}
and similarly for the conjugate.

Alternatively, in the second way, we treat the fields as the independent 
variables, however they occur in $K$, and obtain
\begin{equation}
\delta S = \int \frac{d^4k}{(2 \pi)^4} \left[ \left( \frac{\partial K}{\partial
 \kf} \right)_{T} \delta \kf  + \left( \frac{\partial K}{\partial \kf^*} 
\right)_{T} \delta \kf^* \right]=0 \, , \label{varst}
\end{equation}
where the subscript T denotes a total derivative. Thus the anolog of the 
Euler-Lagrange equations is
\begin{equation}
\left( \frac{\partial K}{\partial \kf} \right)_{T} = 0 \, , 
\label{kwveqt}
\end{equation}
and similarly for the conjugate. This is obviously the same as (\ref{kwveq}).
For the scalar Lagrangian (\ref{kaction}) we vary the action with respect to 
the conjugate field and obtain explicitly
\begin{equation}
\left( \frac{\partial K}{\partial \kf^*} \right)_{T} = (\hbar^2 k^2 - m^2) \kf
= 0 \, , \label{fkg}
\end{equation}
The solution to this may be expressed as
\begin{equation}
\kf = 2 \pi \delta \left( k_0 - \sqrt{\vec{k}^2 + m^2 / \hbar^2} \right) \, 
\tilde{\phi}_+ (\vec{k}) + 2 \pi \delta \left( k_0 + \sqrt{\vec{k}^2 + m^2 / 
\hbar^2}\right) \,  \tilde{\phi}_- (\vec{k}) \label{ksol} \, , 
\end{equation}
where the $\tilde{\phi}_+$ and $\tilde{\phi}_-$ are arbitrary functions of the 
three-vector momentum. This corresponds to a superposition of plane wave 
solutions of the form (\ref{plnwv}).

We may consider the variational principle in wave vector space, in
terms of the function $K$, as the fundamental basis of the theory, and thereby
make no explicit reference to spacetime. This may have some conceptual or 
philosophical merit in that experiments in particle physics are generally
scattering experiments and are more directly related to the momenta of 
particles rather than their position. It may also be possible that such a 
viewpoint might allow easier generalizations of the theory, as we will discuss 
in the next section.

\section{A generalization of the theory}

It has long been a supposed fact of life that the differential equations of 
physics are first or second order. This is well born out by experience in that 
classical mechanics, classical electromagnetism, general relativity, 
nonrelativitstic quantum mechanics, relativistic quantum mechanics, and all the
equations of the standard model of particles are at most of second order. But 
it may well be that we also have developed an unjustified bias in favor of 
second order equations due to mathematical convenience. As such it is 
particularly interesting to consider higher order equations with all their 
inherent dangers and difficulties with boundary conditions. Motivated thereby, 
we consider the following generalization of the scalar wave equation 
\cite{itz},
\begin{equation}
\hbar^2( \partial^2 - a l_{p}^{2} \partial^4 + \dots ) \f + m^2 \f =0 \, , \ \ 
\partial^2 \equiv \ue \frac{\partial^2}{\partial x^{\mu} \partial x^{\nu}} \, ,
\ \ \partial^4 \equiv \partial^2 \partial^2 \, ,  \ \ \text{etc.} \label{gstkg}
\end{equation}
Here  $l_p$ is a characteristic distance, presumably of the order of the Planck
distance, and  $a$ is a dimensionless parameter of order 1, with either sign.
The series represents any reasonable function with a power series
expansion.  This equation does not fit easily into the action formalism in
spacetime since it may be of infinite order in derivatives. If the series
is infinite this equation is, in a sense, nonlocal.

We can easily find plane wave solutions of the form (\ref{plnwv}). These
are solutions if
\begin{equation}
-\hbar^2 k^2 (1 + a l_{p}^{2} k^2 + \dots) + m^2 \equiv -\hbar^2 k^2 f^2(k^2) +
m^2 =0 \label{gdisper} \, .
\end{equation}
As before we note that special relativity tells us that $p^2=m^2$, so that
\begin{equation}
\hbar^2 k^2 f^2(k^2) = p^2 \label{gdisper2} \, .
\end{equation}
Since $f^2(k^2)$ is a scalar and the wave vector and the momentum are 
four-vectors the quotient theorem tell us that they are related by a second 
rank tensor,
\begin{equation}
p_{\mu} = t_{\mu} \, ^{\alpha} k_{\alpha} \label{momentum} \, .
\end{equation}
Substitution of this into (\ref{gdisper2}) shows that the tensor $t_{\mu} \, ^{
\alpha}$ must be a scalar multiple of an element of the Lorentz group, or
\begin{equation}
t_{\mu} \, ^{\alpha} = \hbar f(k^2) \Lambda_{\mu} \, ^{\alpha} \, , \ \ \
\Lambda_{\mu} \, ^{\alpha} \ue \Lambda_{\nu} \, ^{\beta} = \eta^{\alpha \beta}
\, . \label{tlorentz}
\end{equation}
The simplest choice for $\Lambda_{\mu} \, ^{\alpha}$, and the only one that 
does not depend on some intrisic parameter, is the identity $\delta_{\mu}^{
\alpha}$. We accordingly assume, in direct analogy with the assumption in 
section II, a \underline{generalized de Broglie relation}
\begin{equation}
p_{\mu} = \hbar f(k^2) k_{\mu} \, . \label{gendebrog}
\end{equation}
This implies in particular that the wavelength of the particle is given by
\begin{equation}
\lambda = \frac{2 \pi \hbar}{p} f(k^2) \label{genwvlngth}  \, .
\end{equation}
Such a modification may not be easy to test experimentally since the
wavelengths of particles in scattering experiments are not generally
observed, and at sufficiently high energies may not even be observable in
principle.

It is amusing to note that the above generalization of the theory
may be viewed as allowing Planck's constant to become a function of the
wave vector squared, or equivalently the mass of the associated free particle. 
It is for this reason that we retain it explicitly as a bookkeeping tool in all
equations.

Our consideration of the generalized scalar wave equation is not
difficult, but is slightly cumbersome in the spacetime Lagrangian
formalism. It amounts to making the substitution $ \hbar \rightarrow \hbar
f(-\partial^2)$. In wave vector space however it is extremely simple and 
natural. We may view the generalization as multiplying the wave vector by $f(
k^2)$ or equivalently as allowing Planck's constant to be a scalar function 
via $\hbar \rightarrow \hbar f(k^2)$; in either view we substitue $\hbar k_{
\mu} \rightarrow \hbar k_{\mu} f(k^2)$, so the function $K$ in (\ref{kaction}) 
is modified to
\begin{equation}
K = \kf^* [\hbar^2 f^2(k^2) k^2 - m^2] \kf \label{genk} \, .
\end{equation}
The wave equation is then
\begin{equation}
[\hbar^2 f^2(k^2) k^2 - m^2] \kf =0 \label{genwveq} \, ,
\end{equation}
and the solution is
\begin{equation}
\kf = 2 \pi \delta \left( k_0 - \sqrt{\vec{k}^2 + m^2 / \hbar^2 f^2} \right) 
\, \tilde{\phi}_+ (\vec{k}) + 2 \pi \delta \left( k_0 + \sqrt{\vec{k}^2 + m^2 
/ \hbar^2 f^2} \right) \, \tilde{\phi}_-  (\vec{k}) \label{genksol} \, ,
\end{equation}
which is the obvious generalization of (\ref{ksol}).

One might expect that the generalized de Broglie relation (\ref{gendebrog})
would lead to a change in the dispersion relation for the velocity of a
massive free particle. It can however be easily seen that this is not the
case. From (\ref{gdisper2}) with $p^2=m^2$ we have
\begin{equation}
\hbar^2 f^2(k^2) k^2 = m^2 \Rightarrow k^2=(k^0)^2 - \vec{k}^2 = \text{
constant} \, . \label{gendisper3}
\end{equation}
The group or particle velocity is then given by
\begin{equation}
v_g = \frac{d k^0}{d |\vec{k}|} = \frac{|\vec{k}|}{k^0} = \frac{|\vec{p}|}{p^0}
\, , \label{gv}
\end{equation}
which is independent of the function  $f(k^2)$.

\section{ Generalized commutation relation}

The generalized de Broglie relation (\ref{gendebrog}) leads to a modification 
of the canonical position-momentum commutation  relation of quantum mechanics.
This may be obtained as follows. The commutation relation between the
position and wave vector operators is evident, since the wave vector $k_{
\alpha}$ may be expressed as the derivative $i \partial / \partial x^{\alpha}$
in the position representation, and thus
\begin{equation}
[x^{\mu}, \, k_{\alpha}] = -i \delta^{\mu}_{\alpha} \label{comm1} \, .
\end{equation}
Alternatively, in the wave vector representation, the position operator may
be expressed as $x^{\mu} = -i \partial / \partial k_{\mu}$. Then the 
commutator of $x^{\mu}$ and $p_{\alpha}$ can be written as
\begin{equation}
[x^{\mu}, \, p_{\alpha}] = [-i \frac{\partial}{\partial k_{\mu}}, \, p_{\alpha}
] = -i \frac{\partial p_{\alpha}}{\partial k_{\mu}} \, . \label{comm2}
\end{equation}
We may calculate the indicated derivatives from (\ref{gendebrog}) and obtain
\begin{equation}
[x^{\mu}, \, p_{\alpha}] = -i \hbar \left[ f \delta^{\mu}_{\alpha} + 2 f'
k_{\alpha} k^{\mu} \right] \, \ \ \ f' \equiv \frac{df}{d k^2}  \, .
\label{kcomm}
\end{equation}
To obtain this entirely in terms of momentum we define $g(p^2) \equiv f(k^2)$, 
and use  (\ref{gdisper2}) and (\ref{gendebrog}) to get
\begin{equation}
[x^{\mu}, \, p_{\alpha}] = -i \hbar  \left[g \delta^{\mu}_{\alpha} + \frac{2 g'
p_{\alpha} p ^{\mu}}{1 - 2 g' p^2 / g} \right] \, , \ \ \  g' \equiv 
\frac{d g}{d p^2} \, . \label{gencomm}
\end{equation}

As an explicit example suppose that to lowest order
\begin{equation}
f= 1 + a l_{p}^{2} k^2 \label{exmpl} \, .
\end{equation}
Then the commutation relation takes the simple form, to lowest order,
\begin{equation}
[x^{\mu}, \, p_{\alpha}] = -i \hbar  \left[ \delta^{\mu}_{\alpha} + \frac{a}{
m_{p}^{2}} (p^2 \delta^{\mu}_{\alpha} + 2 p_{\alpha} p^{\mu}) \right] \, , 
\end{equation} \label{exmplcomm}
where $m_p \equiv \hbar / l_p$.

We note parenthetically that the above results may also be obtained
directly in the position representation by using
\begin{equation}
p_{\mu} = i \hbar f(-\partial^2) \frac{\partial}{\partial x ^{\mu}} 
\end{equation}
and assuming a power series representation for $f$, but the derivation is
somewhat more tedious than the above.

\section{The generalized uncertainty principle}

The commutation relation obtained in the previous section leads
immediately to a generalization of the Heisenberg uncertainty principle,
variously called the extended uncertainty principle, or the gravitational
uncertainty principle, or the generalized uncertainty principle (GUP). We
recall that the uncertainty principle of quantum mechanics gives the
product of the variances of two observables (or Hermitian operators) $A$ and
$B$ in terms of their commutator,
\begin{equation}
(\Delta A)^2 (\Delta B)^2 \ge \frac{1}{4} \left| \langle [A,B] \rangle 
\right|^2 \, . \label{rms}
\end{equation}
The variances are for a series of measurements of the observables and
the expectation value is for whatever state is being measured. We apply
this to the position and momentum operators in the $x$ direction and use the
approximate relation (\ref{exmplcomm}) from the preceding section, with $\alpha
= \mu=1$, to obtain
\begin{equation}
\Delta x \Delta p \ge \frac{\hbar}{2} \left( 1+  \frac{am^2}{m_{p}^{2}} - 
\frac{2a}{m_{p}^{2}} \langle p_{x}^{2} \rangle \right) \label{uncxp} \, .
\end{equation}

Consider now an ordinary particle such as an electron with $m \ll m_{p}$ at 
rest, with momentum expectation equal to zero. If it is subjected to a position
measurement localizing its position to a very small region its momentum
spread will be very large. Then, since
\begin{equation}
\Delta p_{x}^{2} = \langle p_{x}^{2} \rangle - \langle p_{x} \rangle^{2} =
\langle p_{x}^{2} \rangle
\end{equation}
equation (\ref{uncxp}) becomes
\begin{equation}
\Delta x \Delta p \ge \frac{\hbar}{2} - \frac{a \hbar}{m_{p}^{2}} \Delta p^2 \,
, \ \ \ \Delta x \ge \frac{\hbar}{2\Delta p} - a l^{2}_{p} 
\frac{\Delta p}{\hbar} \ , . \label{gup}
\end{equation}
This is precisely the GUP, as usually obtained in string theory \cite{ven} or 
by consideration of gravitational effects in a quantum measurement process 
\cite{gup}. Moreover, for agreement with the usual GUP as discussed below, we 
choose $a \simeq -1$.

\section{ Speculation on Gravity}

The GUP is a well-known result of string theory \cite{ven}, and is related to
the interesting physical notions of duality and a minimum length \cite{witten}.
It can be obtained in considerable generality using only basic concepts of 
quantum mechanics and gravity as is done in references \cite{gup} and 
\cite{gup2}. A very brief derivation using dimensional arguments may be made as
follows: consider a photon being used to measure the position of an electron, 
as in the classic discussion of Heisenberg \cite{qm}. In addition to the usual 
quantum mechanical position uncertainty, $\Delta x \simeq \hbar / \Delta p$, 
there must be an uncertainty due to the gravitational field of the photon. The 
extra uncertainty should accordingly be linear in the gravitational constant G 
and in the energy or momentum of the photon. Moreover, The momentum of the 
photon should be of the order the uncertainty in the electron momentum, so the 
extra term should be proportional to $G \Delta p$. Thus the GUP must be, on 
dimensional grounds,
\begin{equation}
\Delta x \simeq \frac{\hbar}{\Delta p} + l_{p}^{2} \frac{\Delta p}{\hbar} \, .
\label{gup2}
\end{equation}
This is in qualitative agreement with (\ref{gup}).

In the derivation of the GUP in section VI we made no mention of gravity, but 
have obtained the result form the generalized de Broglie relation and the 
consequent commutation relation. We are led to speculate that the 
generalization may in some way generate gravity. In this view gravity may not 
be a fundamental force, but instead, it may be generated by quantum
effects. This is not unreasonable physically: general relativistic gravity
is characterized by its breaking of translational invariance in spacetime,
and in a similar way a nonconstant function $f(k^2)$  breaks the translational
invariance of the action in wave vector space.

A suggestion that is similar in spirit was made long ago by Sakharov
\cite{sak}, who speculated that gravity may not be fundamental, but induced by 
the residual effects of fundamental quantum fields on the vacuum, with the 
Lagrangian playing the role an elastic stress. Such a view of gravity
has the merit of providing some understanding of why gravity is so weak. In 
the Sakharov view gravity is weak because the Planck scale provides an energy 
momentum cutoff that is very large. In the present view gravity is weak because
the relation between momentum and wave number differs from linear only for very
large momenta.

\section{ Interactions }

Although the main parts of this paper deal with a free scalar field, it
should be emphasized that interactions are easily accomadated in the
context of wave vector space. As a simple example of this we consider an
interaction in the spacetime Lagrangian of the form
\begin{equation}
L_I = g \f^* \f \Omega(x) \, , \label{inter}
\end{equation}
where $\Omega$ is a given external field and $g$ is a coupling constant. The 
Lagrangian function in wave vector space is then modified from (\ref{genk}) to
\begin{equation}
K = \kf^* [\hbar^2 f^2(k^2) k^2 - m^2] \kf - g \int \frac{d^4 l}{(2 \pi)^4}
\tilde{\phi}(k+l)^* \kf \tilde{\Omega}(l) \, . \label{kinter}
\end{equation}
That is, the point interaction term becomes a convolution, so the
interaction is nonlocal in wave vector space. The extremum condition, 
(\ref{kwveq}) or (\ref{kwveqt}), then is an integral equation
\begin{equation}
 [\hbar^2 f(k^2) k^2 - m^2] \kf = g \int \frac{d^4 l}{(2 \pi)^4} \tilde{\phi}(k
-l) \tilde{\Omega}(l) \, , \label{kwveqint}
\end{equation}
where we have made a shift of $l$ in the wave vector space. This is easily
solved by a perturbation expansion of $\kf$ of the form
\begin{equation}
\kf = \tilde{\phi}_0 (k) + g \tilde{\phi}_1 (k) + \dots = \sum_{n} g^n 
\tilde{\phi}_n (k) \ , . \label{gexp}
\end{equation}
We substitute this into (\ref{kwveqint}) and get a set of iterative equations
\begin{equation}
[\hbar^2 f(k^2) k^2 - m^2] \tilde{\phi}_n (k)  = g \int \frac{d^4 l}{(2 \pi)^4}
\tilde{\phi}_{n-1} (k-l) \tilde{\Omega}(l) \, . \label{itertveq}
\end{equation}
As the zeroth order or unperturbed solution we naturally choose an eigenstate 
of the wave vector $k_i$,
\begin{equation}
\tilde{\phi}_0 (k) = (2 \pi )^4 \delta(k - k_i ) \label{zero} \, ,
\end{equation}
and obtain the series solution
$$
\kf = (2 \pi )^4 \delta(k - k_i ) +g \frac{1}{\hbar^2 f(k^2) k^2 - m^2} \tilde{
\Omega}(k - k_i) \ + 
$$
\begin{equation}
g^2 \frac{1}{\hbar^2 f(k^2) k^2 - m^2} \int \frac{d^4 l}{(2 
\pi)^4} \tilde{\Omega}(l) \frac{1}{\hbar^2 f[(k-l)^2] (k-l)^2 - m^2} \tilde{
\Omega}(k - k_i-l) + \dots \label{sol} 
\end{equation}
This we recognize as the usual Feynman expansion, but with a modified 
propagator given by 
\begin{equation}
D(k^2) = \frac{1}{\hbar^2 f(k^2) k^2 - m^2}  \label{propgtr} \, .
\end{equation}

As an illustrative example let us take the function $f$ to be linear, as in 
(\ref{exmpl}), but expressed in terms of the Planck mass instead of the Planck
length:
\begin{equation}
f(k^2) = 1 - l_{p}^{2} k^2 = 1 - \frac{\hbar^2}{m_{p}^{2}} k^2 \, .
\label{lnrf}
\end{equation}
We note parenthetically that the momentum of the particle will become zero when
$k^2 = m^{2}_{p} / \hbar^2$. Massless particles could a null wave vector or
a wave vector with magnitude $m_{p} / \hbar$, the inverse of the Planck length!
Combining equations (\ref{propgtr}) and (\ref{lnrf}), we obtain the propagator
\begin{equation}
D(k^2) = \frac{1}{\hbar^2 k^2 - \hbar^4 k^4 / m_{p}^{2} - m^2}  \, .
\label{propgtr2}
\end{equation}
It has poles at
\begin{equation}
\hbar^2 k^2 = \frac{m_{p}^{2}}{2} \left( 1 \pm \sqrt{1 + 4 m^2 /m_{p}^{2}} 
\right) \simeq \left \{ \begin{array}{rr} m^2 \text{ for } - \\
m_{p}^{2} \text{ for } + \end{array} \, , \right. \label{poles}
\end{equation}
where the approximation is valid for $m^2 \ll m_{p}^{2}$. In this approximation
the propagator factors to
\begin{equation}
D(k^2) = \frac{1}{\hbar^2 k^2 - \hbar^4 k^4 / m_{p}^{2} - m^2} \simeq \left(
\frac{1}{1 - \hbar^2 k^2 / m_{p}^{2}} \right) \left( \frac{1}{\hbar^2 k^2 -
 m^2} \right) \, . \label{facprop}
\end{equation}
That is, a form factor factor appears as multiplying the usual
propagator. Such a form factor is usually associated with a modification of the
vertex in quantum field theories, but here it appears as a result of the
modified de Broglie relation.

Further discussion of quantum field theories and perturbation theory, in 
particular gauge theories, will follow in another paper.

\section{ Summary}

We have generalized the de Broglie relation (\ref{momtm}) to the four-vector
relation (\ref{gendebrog}), which involves an arbitrary scalar function. Thus 
the wavelength of a particle of a given momentum depends on its invariant mass.
This was done in the context of the Klein-Gordon or scalar wave equation. In 
spacetime the generalized equation contains an arbitrarily large number of 
derivatives; in a formalism involving a Lagrangian type function in wave vector
space the derivation is simpler and more natural. We emphasize that the wave 
vector space formalism is more closely related to actual laboratory experiments
in high energy scattering, and that spacetime on a small scale need
not even exist in this point of view.

As a result of the generalized de Broglie relation the canonical commutation 
relations are modified, and lead to a generalized uncertainty principle (GUP). 
Since the GUP may otherwise be obtained from a consideration of gravitational 
effects in the quantum measurement process the question naturally arises as to 
whether gravity may in some way be related to, or generated by, quantum 
effects, and not be an independent fundamental force.

Interactions are easily accomodated in the wave vector space
formalism. A Feynman type expansion for the scalar field with a modified
propagator was given as an example. Further studies will deal with
perturbative field theories, especially with gauge theories such as QED.

\section*{Acknowledgements}

This work was supported by NASA grant NAS 8-39225 to Gravity Probe B. Finally 
the Gravity Probe B theory group at Stanford provided many stimulating 
discussions.

\end{document}